\documentclass{llncs}
\usepackage{amssymb}
\usepackage{version}

\title{Interest Prohibition and \\ Financial Product Innovation}

\author{J.A. Bergstra \and C.A. Middelburg}
\institute{Section Theory of Computation, Informatics Institute,
           University of Amsterdam \\
           Science Park~904, 1098~XH Amsterdam, the Netherlands \\
           \email{J.A.Bergstra@uva.nl, C.A.Middelburg@uva.nl}
          }

\begin{document}

\maketitle

\begin{abstract}
%% 59  %%
We give a rough sketch of the Judaic, Greek, Islamic and Christian
positions in the matter of interest prohibition during the last few
millennia and discuss the way in which interest prohibition is dealt
with in Islamic finance, the problems with authority-based arguments for
interest prohibition, and the prospects of interest prohibition with the
advent of electronic money.
\par\addvspace{1.5ex} \small {\sl Keywords:}
interest prohibition, Islamic finance, authority-based argument,
electronic money.
\nolinebreak
\end{abstract}

\section{Introduction}
\label{sect-intro}

Many different words in different languages have been used for
interest.
In particular, usury went from denoting all forms of interest to
denoting excessively high interest rates.
Preventing or forbidding the occurrence of usury in the latter sense has
become standard practice in the Western world.
Forbidding the occurrence of usury in the former sense underlies Islamic
finance.

Some Islamic authors indicate that loans at interest can bring people in
severe problems because they have no way to redeem their debt (see
e.g.~\cite{Sid05a}).
The heroic battle fought by many countries against their excessively
large and interest bearing government debts is an additional reason to
consider the phenomenon of interest with caution and perhaps to look for
its replacement by an alternative to it.
Difficulties caused by the charging of interest on loans at an
individual level are convincingly described in~\cite{Lew99a}.

In non-Islamic countries, the debate about the status of interest has
disappeared and opposition to interest is seen as a matter that has
been overcome in a process of progress.
Papers about the subject have invariably a historic character, and the
suggestion that old objections against interest are still of some value
seems very remote.
One might even say that conventional finance is based on interest,
anyhow it is interest permissive.
A wealth of information concerning the history of views on interest
prohibition in the non-Islamic world has been published (see
e.g.~\cite{Ack81a,Gor82a,VM98a,Sen06a}).

Islamic finance is an approach to finance and banking which has not
forgotten old objections to the interest mechanism.
Islamic finance is not interest permissive.
The remarkable success of systems of Islamic finance has been
documented in many sources (see e.g.~\cite{DP99a,ElG01a}).
An account of the different histories of interest restrictions in the
Christian world and the Islamic world can be found in~\cite{Rub07a}.

The rest of this paper concerns several topics relevant to interest
prohibition and Islamic finance.
By way of introduction, we first give an outline of the discussions
about the pros and cons of interest that have been going on for almost
three millennia and a brief account of the main policies for dealing
with interest.
Next, we give a constructive perspective on Islamic finance and
present some interesting features of this perspective.
One of these features is that the hard to verify claim that Islam
forbids all forms of interest need not be accepted as a basic rule.
We also have a close look this claim.
After that, we discuss the prospects of interest prohibition in the
light of the advent of electronic money and mention some weak points of
current Islamic finance.

\section{Outline of the History of Interest}
\label{sect-history}

Discussions about the merits and demerits of interest have been going on
for almost 3000 years.
Around 620~BC, it was proclaimed that Jews were not allowed to impose
payment of interest to other Jews but that they were allowed when
dealing with non-Jews (see the Old Testament, Book of Deuteronomy
23:19--20).
However, the earliest Judaic recommendation against interest taking
dates from 800~BC or earlier (see the Old Testament, Book of Exodus
22:25--26).
In addition, irredeemable debts were considered undesirable in the
Judaic tradition.
It was therefore proclaimed that at regular intervals debts should be
remitted, allowing debtors to proceed their lives without the burden of
previous debts (see the Old Testament, Book of Deuteronomy 15:1--2).
Ancient Jews also preferred interest-free loans over gifts, as a gift
suggests that its receiver has reached the stage of a beggar.

The Greek philosopher Aristotle (384~BC--322~BC) condemns in his
``Politics''\linebreak[2] (Book~I, Parts~9--10) both getting money by
lending money at interest and getting money by first buying goods and
then selling them for a greater sum because the end of these forms of
exchange is solely accumulation of wealth.
He also states that getting money by lending money at interest is worse
because, amounting to the birth of money from money, it is the most
unnatural way of getting wealth.
The Italian philosopher and theologian Thomas Aquinas (AD~1225--AD~1274)
argues in his ``Summa Theologiae'' (Part~SS, Question~78) that paying
interest for a loan together with returning the principal amount implies
a double payment for the same thing.
The statements of both philosophers suggest a confusion between two
forms of use: instrumental use that leaves the instrument ready (which
may justify interest or rather rent) for the next round and consumptive
use which justifies the restitution of costs (principal amount).

Thomas Aquinas also adopted Aristotle's judgement that interest is
unnatural and for that reason should not be allowed, and so did many
other Christian authors.
This brought the Church of Rome in the sixteenth century to a formal
condemnation of the interest mechanism by the council of Trento.
This condemnation still stands, although the practice of paying interest
has been reluctantly accepted in the encyclical ``Rerum Novarum'' issued
by the Pope Leo~XIII in AD~1891.

In the Qur'an, the prophet Muhammad (AD~570--AD~632) explicitly
indicated the so-called doubling debt scenario as being against God's
will (Verse~3:130), though without suggesting the need of earthly
punishments for those who claim or receive such an excessive interest.
However, he predicted serious posthumous consequences.
Subsequent authors, notably Abu Bakr al-Jassas (died AD~981), have
argued that the indications revealed through Muhammad must be understood
as special cases of more general requirements, in particular the
requirement that interest is never applied, which according to him also
covers the special case of the doubling debt scenario.

\section{Policies for Dealing with Interest}
\label{sect-policies}

We can distinguish three different policies for dealing with interest if
these are deemed problematic: interest moderation, interest prevention,
and interest prohibition.
\begin{description}
\item
\emph{Interest moderation.}
Moderation involves the imposition of upper bounds to interest rates and
to the size of loans, both essential to avoid that people destroy their
lives by accepting too large debts.
Interest moderation is applied in all systems of conventional finance.
The upper bounds to interest rates are relative to the financial
capabilities of the borrower.
\footnote
{It is remarkable that interest moderation is nowadays also applied at
 the level of national states where states issue guarantees so that
 states in severe debt can borrow new money at acceptable interest
 rates.
}
\item
\emph{Interest prevention.}
Prevention involves redesign of financial methods so that loans play a
lesser role.
Debt prevention is the main means for achieving interest prevention.
Necessary debts will still carry interest.
Necessary debts can occur within the system if the needs of parties
cannot be satisfied in another way.
\item
\emph{Interest prohibition.}
Interest prohibition comes on top of interest prevention: interest
prohibition excludes all loans between parties who subscribe to the
prevention policy.
Only when deals must be made with parties not adhering to the interest
prohibition policy, it is acceptable that a party who adheres to it will
either pay or receive interest.
\end{description}

Islamic finance has been designed in such a way that it implements both
interest prevention and interest prohibition, and it does so to the
extent that there is no need left for interest moderation.

\section{A Constructive Perspective on Islamic Finance}
\label{sect-constr-persp}

It has a significant disadvantage to explain Islamic finance as a
financial system that implements interest prevention and interest
prohibition to the extent that all remaining loans are interest-free.

First of all, Islamic finance aims at further restrictions and it aims at
additional and more important objectives.
Explanations of these additional restrictions and the key objectives are
given in~\cite{ZH01a,ElG06a,GW09a,Iqb09a}.
More importantly, Islamic finance is best understood as being under a
steady development, where its product portfolio grows under the
constraint that each product on offer is approved by a significant
majority of Islamic scholars.
Their approval concerns compliance with requirements that stem from
Islam.
Different scholars may have different views on the matter, but a product
will be put on the market by most institutions of Islamic banking only
when a majority view has been reached.

Using this perspective, interest prohibition and interest prevention
serve as heuristics that help financial engineers to predict the
judgement of scholars (usually cooperating in Shari'ah boards).
If interest bearing debt normally occurs during the lifetime of a
financial product, it is quite likely that scholars will not approve of
it.
And if they do so initially, they may change their mind in a later stage
once the occurrence of interest payments during the term of a product
has becomes clear.
This problem may be spotted even in a fraction of the activities to
which a product gives rise.
So-called concealed interest is only found when scrutinizing the
progression of actions to which the application of a product in actual
circumstances may lead.

We call this viewpoint on Islamic finance a constructive perspective
because the bottom-up construction of a product portfolio together with
the practices for its adequate use is highlighted.
This perspective has the following interesting features:
\begin{itemize}
\item
Precise definitions of what constitutes interest, and to which forms of
debt it applies, do not matter.
It is left to scholars to make up their minds about these matters.
Approved products are considered valid, and because the screening is
quite strict that judgement will stand for many years to come.%
\footnote
{In~\cite{BM10b}, we have given a detailed specification of a savings
 account.
 It appears that it is unexpectedly complex and that many options are
 left when it must be decided what exactly is to be forbidden once the
 savings account is labelled as interest bearing.
}
\item
The question how to prove from religious sources that interest is
forbidden is of marginal importance only.
Different scholars may have different views on the matter.%
\footnote
{Subhani~\cite{Sub01a} states that, if it is agreed that Islam disallows
 interest bearing debts, some rationale for the disallowance needs to be
 presented.
 In the form in which Subhani states the problem, it looks quite
 intractable.
}
Hard to verify claims such as ``Islam forbids all forms of interest''
need not be taken as axioms.
\item
If views concerning interest become more permissive, the only effect
will be that additional products are now allowed.
There is no inconsistency that arises from such a change of opinion.
\item
All financial products proposed within Islamic finance can be considered
valid products on the non-Islamic financial market as well.%
\footnote
{In~\cite{BM10b}, we have coined the phrase ``Reduced Product Set
 Finance'' (RPSF) as a generic name for designs of financial systems that
 make use of a reduced set of products.
 Islamic finance is a prime example of an RPSF.
}
\item
Interest prevention and interest prohibition function as informal
drivers of financial product innovation.
The family of new products that is brought on the market by institutions
supporting Islamic finance often allows for their clients to operate
with less risk than they would have been exposed to in the case where
they had made use of the product family of conventional banks.
\item
The new products are accessible for non-Islamic clients, and these
may be attracted by the strongly risk-aversive policy that underlies
these products.
Product innovation as pursued by Islamic banking institutions can help
them to perform competitively in a market that allows for other banking
forms~too.
\end{itemize}

\section{On Authority-Based Arguments for Interest~Prohibition}
\label{sect-authority}

Observers from outside Islam (including the authors) have enormous
difficulties with understanding hard to verify authority-based arguments
for interest prohibition where it is not clear how the authority has
come about.
The claim that Islam forbids all forms of interest is an important
example of such an argument.
It is essentially the claim made by Abu Bakr al-Jassas around AD~980.
The argument is seldom used by contemporary Islamic scholars, but it is
used in canvassing texts (see e.g.\ \cite{Ismail}).
Authority based arguments have their own logic (see e.g.~\cite{Wal08a})
and that is a complex matter.
They descend in historic chains with authoritative persons, groups,
events, or texts as sources.
A closer look at the claim that Islam forbids all forms of interest
reveals the following:
\begin{itemize}
\item
The literature indicates that most Islamic scholars who base their
judgement primarily on reading of the Qur'an accept that Islam prohibits
the doubling debt scenario, but do not agree that prohibition goes much
further (see e.g.~\cite{Far07a,Rah64a}).
\item
Most authors who state or repeat the claim do so without further
argument.
That is a nearly circular matter given the fact that these authors on
their own do not generate the authority to produce the claim as a source
and no reference is made to previous sources of higher authority either.
\item
In principle, it cannot be excluded that many sources state the claim
while in fact none of these sources does so in a way that by itself
carries enough weight for the claim to be accepted initially.
\item
Non-Islamic literature about revealed sources indicates that basing a
claim on such sources needs to be done with care.
Sources develop the status of having been revealed because of the
authority that is afterward assigned to their contents, instead of the
other way around.
By scrutinizing the concept, it is demonstrated in~\cite{Whi10a} that
revealed status of sources is not an a priori matter which can be
asserted independently of their contents.
\item
A number of authors point at the advantages of interest prevention and
interest prohibition as arguments in favour of the design of Islamic
financial systems.
They create confusion about the degree of authority that must be
attributed to the claim.
If its authority is unquestioned, then remarks about the positive
consequences of its implementation are nice but cannot replace the
appeal to the authoritative claim.
\item
Some authors argue that interest does not occur in an Islamic society
and that from that empirical fact a design rule can be inferred which is
plausibly phrased as the claim.
If this is true, the empirical fact needs an explanation first and it is
not obvious at first sight that the explanation is found in the prior
existence or awareness of the claim in an implicit form.
\item
Even if the claim has been established with sufficient authority at some
stage, why must it be considered valid for money in the forms that it
takes hundreds or thousands of years later.
It is very plausible that the concept of money goes through an evolution
which puts each conceptual analysis in question.
The latest authority who wrote on interest with impact was Thomas
Aquinas --- authoritative Islamic writers preceded him.
When he wrote on interest, money and interest was not thought of in
terms of inflation, deflation, fiat money, liquidity preference,
equilibrium on a financial market et cetera.
\end{itemize}

\section{Evolving Money}
\label{sect-evolving-money}

Although money seems to have been of a constant quality from the seventh
century BC to the eighteenth century AD, changes are now becoming more
likely.
From the eighteenth century, fiat money has become dominant, but its
presence in the form of coins or banknotes has become increasingly
marginal.
We are now looking into a future which is dominated by electronic money.
As we write in~\cite{BM06e}, money will become a computational
phenomenon.
Its creation will be physically trivial, but it will be blocked by
secure computation technology.
It is not unlikely that the familiar property that money can be applied
for all purposes by every owner will disappear.
The purchasing power of money is likely to become dependent on the
social and financial position of its owners, whether we like it or not.

It is now becoming easy to devise a financial system in which money is
always tagged to the electronic identity of its owner.
By requiring that money can only be spent by its owner, it is easy to
prevent lending in such a system.
In that way, debts and interest disappear.
We think that, now that money is becoming electronic, old intuitions
cease to prevail.
All features of its use can be switched on and off on a personal basis.
It is technically possible to allow debts only in special circumstances
and to disallow interest payment fully.
This implies that the heuristic value of interest prohibition will
diminish and that the question how one expects the financial system to
work at large to become more prominent.

These considerations lead to the observation that interest prohibition
derives its significance from the fact that those who intend to comply
with it expect to make use of the same money as those who do not have
that intention.
An interest prohibitive community and an interest permissive community
make deliberately use of the same money, which is a means of
communication between them.
If it is true that the intention to share money is predominant for both
communities, they will have to interact when further innovations of the
financial system are involved.
If this is right, both communities have an interest in understanding the
details of a common interest prohibition theory (IPT, see~\cite{BM11a}),
because they must be able to talk about their disagreements in technical
detail.
They will also need some systematic embedding of the reasoning methods
that are employed.
These matters can be laid down in an application specific informal logic
(ASIL, see again~\cite{BM11a})).

\section{Critique of Current Islamic Finance}
\label{sect-critique}

In spite of its remarkable progress, critical remarks have been put
forward concerning the present state of Islamic finance.
For instance, adhering to interest prohibition may fail to lead to a
financial system compatible with Islamic objectives at large (see
e.g.~\cite{ElD97a}).
It has been stated by many authors that international coordination
is needed to strengthen the system of Islamic finance, which at
closer inspection consists of a patchwork of systems of Islamic
finance each with slightly different methods and rules.
In~\cite{Ber11a}, the point is made that centralization may also
constitute a significant risk for the quality of Islamic finance as a
tool for promoting other Islamic objectives.

From~\cite{BM10b}, we take the following explanation of what seems to be
the greatest weakness of interest prohibition: loans will incur far
higher transaction costs than interest compensated loans if focus is on
profit-loss sharing --- the key principle of financial product design in
Islamic finance.
If $X$, rather than borrowing $p$ to $Y$, provides $p$ to $Y$ for
combining forces in a joint enterprise that is mainly managed by $Y$,
whereas $X$ is entitled to some fraction of its revenues, $X$ inevitably
is dependent on being correctly informed about $Y$'s results.
Providing this detailed business information is quite expensive.
Even worse, $X$ and $Y$ seem to have conflicting interests, as $Y$ may
prefer to underestimate his profits from the joint enterprise so that he
needs to channel less revenues to his partner $X$.
The simplification introduced by a loan with interest, with respect to a
profit-loss sharing participation in another party's enterprise, is that
a complete correspondence of objectives between both parties is obtained
if $X$ is not fraudulent (otherwise both options are unsatisfactory
anyhow).
It follows that, in the case of a loan with interest, $X$ can manage his
part of the contract on the basis of far more abstract information about
$Y$.
The information in question may be obtained in a less intrusive fashion,
which is profitable for maintaining proper relations between $X$ and
$Y$.
Of course, this argument would disappear if the prohibition of interest
was judged on a case by case basis.
However, such flexibility is absent in recent writings on interest
prohibition.

\section{Concluding Remarks}
\label{sect-conclusions}

With money becoming more flexible as a consequence of non-stoppable
computerization, more sophisticated patterns of use of money can be
built into the system.
Preferences such as interest prohibition as well as the large family of
financial products that have been developed within Islamic finance to
date will become more important in the age of exclusively electronic
money that lies ahead.
Although interest permission and interest prohibition seem to impose
incompatible design requirements for financial systems, exactly that
complexity may drive the development of new generations of financial
products which primarily aim at obtaining system stability and
reliability instead of accommodating an almost unlimited growth which
has been driving ``conventional'' financial product development.

\bibliographystyle{splncs03}
\bibliography{IF}

\begin{comment}
If money is gold, payment of interest by a borrower requires the
borrower to create gold which is a capacity that a human being ought not
even try to acquire.
In this case, asking for interest payment is wrong, assuming by default
that the borrower is not exploiting his personal gold mine (see also the
\emph{ahadith} dealing with the prohibition of interest quoted
in~\cite[page~50]{Sub01a}).
\end{comment}

\end{document}